\pdfoutput=1
\documentclass[10pt,a4paper,aps,showpacs,pdftex,prl,twocolumn,preprintnumbers,superscriptaddress]{revtex4-1}

\usepackage{amsfonts,amsmath,amssymb}
\usepackage{graphicx}
\usepackage[per-mode=symbol]{siunitx}
\usepackage{hyperref}

\graphicspath{{./}}
\graphicspath{{./Figures/}}

\DeclareSIUnit\intensity{\watt\per\centi\meter\squared}
\DeclareSIUnit\fieldstrength{\volt\per\centi\meter}

\newcommand{\degree}{\ensuremath{^\circ}}%
\newcommand{\Estat}{\ensuremath{\textbf{E}_{\textup{stat}}}}%
\newcommand{\Estatabs}{\ensuremath{\text{E}_{\textup{stat}}}}%
\newcommand{\Ealign}{\ensuremath{\textbf{E}_{\textup{align}}}}%
\newcommand{\ie}{i.\,e.}%
\newcommand{\Ialign}{\textup{I}_\textup{align}}%
\newcommand{\Nuptotal}{\ensuremath{\text{N}_\text{up}/\text{N}_\text{tot}}}%
\newcommand{\dstate}[2]{\ensuremath{\left|\tilde{#1},\tilde{#2}\right>}}%
\newcommand{\pstate}[3]{\ensuremath{\left|#1,#2,#3\right>}}%
\newcommand{\ppstate}[3]{\ensuremath{\left|#1,#2,#3\right>_\text{p}}}%

\newlength{\figwidth}
\newlength{\figwidthsmall}
\begin{document}
\title{Making the best of mixed-field orientation of polar molecules: A recipe for achieving adiabatic dynamics in an electrostatic field combined with laser pulses}%
\author{Jens H. Nielsen}%
\affiliation{Department of Physics and Astronomy, Aarhus University, 8000 Aarhus C, Denmark}
\author{Henrik Stapelfeldt}%
\email[Corresponding author: ]{henriks@chem.au.dk}%
\affiliation{Department of Chemistry, Aarhus University, 8000 Aarhus C, Denmark}
\affiliation{Interdisciplinary Nanoscience Center (iNANO), Aarhus University, 8000 Aarhus C,
   Denmark}%
\author{Jochen K\"upper}%
\email[Corresponding author: ]{jochen.kuepper@cfel.de}%
\affiliation{Center for Free-Electron Laser Science, DESY, 22607 Hamburg, Germany}%
\affiliation{Department of Physics, University of Hamburg, 22761 Hamburg, Germany}%
\affiliation{Fritz-Haber-Institut der Max-Planck-Gesellschaft, 14195 Berlin, Germany}%
\author{Bretislav Friedrich}%
\affiliation{Fritz-Haber-Institut der Max-Planck-Gesellschaft, 14195 Berlin, Germany}%
\author{Juan J.\ Omiste}%
\author{Rosario Gonz\'{a}lez-F\'{e}rez}%
\email[Corresponding author: ]{rogonzal@ugr.es}%
\affiliation{Instituto Carlos I de F\'{\i}sica Te\'orica y Computacional and Departamento de
   F\'{\i}sica At\'omica, Molecular y Nuclear, Universidad de Granada, 18071 Granada, Spain}
\date{\today}
\begin{abstract}\noindent%
   We have experimentally and theoretically investigated the mixed-field orientation of
   rotational-state-selected OCS molecules and we achieve strong degrees of alignment and
   orientation. The applied moderately intense nanosecond laser pulses are long enough to
   adiabatically \emph{align} molecules. However, in combination with a weak dc electric field, the
   same laser pulses result in nonadiabatic dynamics in the mixed-field \emph{orientation}. These
   observations are fully explained by calculations employing, both, adiabatic and non-adiabatic
   (time-dependent) models.
\end{abstract}
\maketitle

Creating oriented samples of polar molecules, \ie, molecules with their dipole moment preferentially
pointing towards one hemisphere rather than the opposite, has been a long standing goal in molecular
sciences. It was originally motivated by the crucial role played by orientation in chemical reaction
dynamics~\cite{Brooks1976}. More recently, its importance in novel applications such as
(fs-time-resolved) photoelectron angular distributions~\cite{Holmegaard:NatPhys6:428,
   Bisgaard:Science323:1464, Hansen:PRL106:073001},
diffraction-from-within~\cite{Landers:PRL87:013002}, or high-order harmonic generation~\cite{NRC:NO}
has been realized.

Early methods exploited purely electrostatic fields. Using an electric multipole focuser, molecules
in a single low-field-seeking quantum state can be selected due to their first-order Stark
effect~\cite{Gordon:PR99:1264, Reuss:StateSelection, Rakitzis:Science303:1852}. The degree of
orientation is determined, and also limited, by the selected state. Alternatively, a strong
homogeneous electric field can create so-called brute-force orientation~\cite{Loesch:JCP93:4779,
   Friedrich:Nature353:412}. This method requires very high electric field strengths and works best
for rotationally cold molecules with large permanent dipole moments.

In 1999 a method based on the combined action of a moderately intense, nonresonant laser field and
an electrostatic field was proposed~\cite{friedrich_enhanced_1999}. For the case that the laser
field is turned on significantly more slowly than the rotational period(s) of the molecule adiabatic
behavior was assumed. The time-independent calculations showed that the degree of orientation could
be nearly perfect under conditions present in many experimental setups. Furthermore, the degree of
alignment, \ie, the confinement of the molecular axes to space-fixed axes, could also be very high.
In addition, the method should be generally applicable to a broad range of molecules and, therefore,
promises the availability of strongly oriented and aligned molecules for various applications.
Experiments performed in the first half of the 2000s showed the feasibility of the method but the
degree of orientation observed was moderate~\cite{Buck:IRPC25:583, Sakai:PRL90:083001}. A major
reason for the weak orientation was that while the individual pendular states are strongly oriented,
these states arise in pairs whose members are oriented oppositely with respect to one another.
Consequently, the resulting overall degree of orientation, obtained as the weighted average over the
populated quantum states, diminishes compared to what is expected for very cold or even single-state
molecular ensembles. A significant improvement in the experimental capabilities was reported in 2009
when quantum-state selected molecules were employed as targets leading to strongly enhanced
orientation~\cite{Holmegaard:PRL102:023001, ghafur_impulsive_2009, filsinger_quantum-state_2009}.
However, it was already realized that an adiabatic description is not sufficient to reproduce the
experimental observations~\cite{omiste-pccp-2011}.

In the present work, we seek the maximum of achievable orientation, as predicted by the original
adiabatic description. Therefore, we prepare a nearly pure ($92^{+3}_{-5}$~\%)
rotational-ground-state ensemble of OCS molecules~\cite{Nielsen:PCCP13:18971} and use a laser pulse
that is sufficiently strong to ensure sharp alignment and that is turned on a 100 times slower than
the rotational period of the molecules. Our experimental observations are, however, at odds, both
qualitatively and quantitatively, with the predictions of the original
theory~\cite{friedrich_enhanced_1999}. Instead, the experimental findings, exploring the dependence
of the orientation on both the laser intensity and on the static field strength, can be rationalized
by solving the time-dependent Schr\"odinger equation describing the mixed-field orientation. Our
analysis directly shows the nonadiabatic coupling of the two sublevels of the near-degenerate
doublets, created by the laser field, and allows us to predict the experimental conditions needed to
ensure adiabatic dynamics.

\begin{figure}
   \centering
   \includegraphics[width=\figwidth]{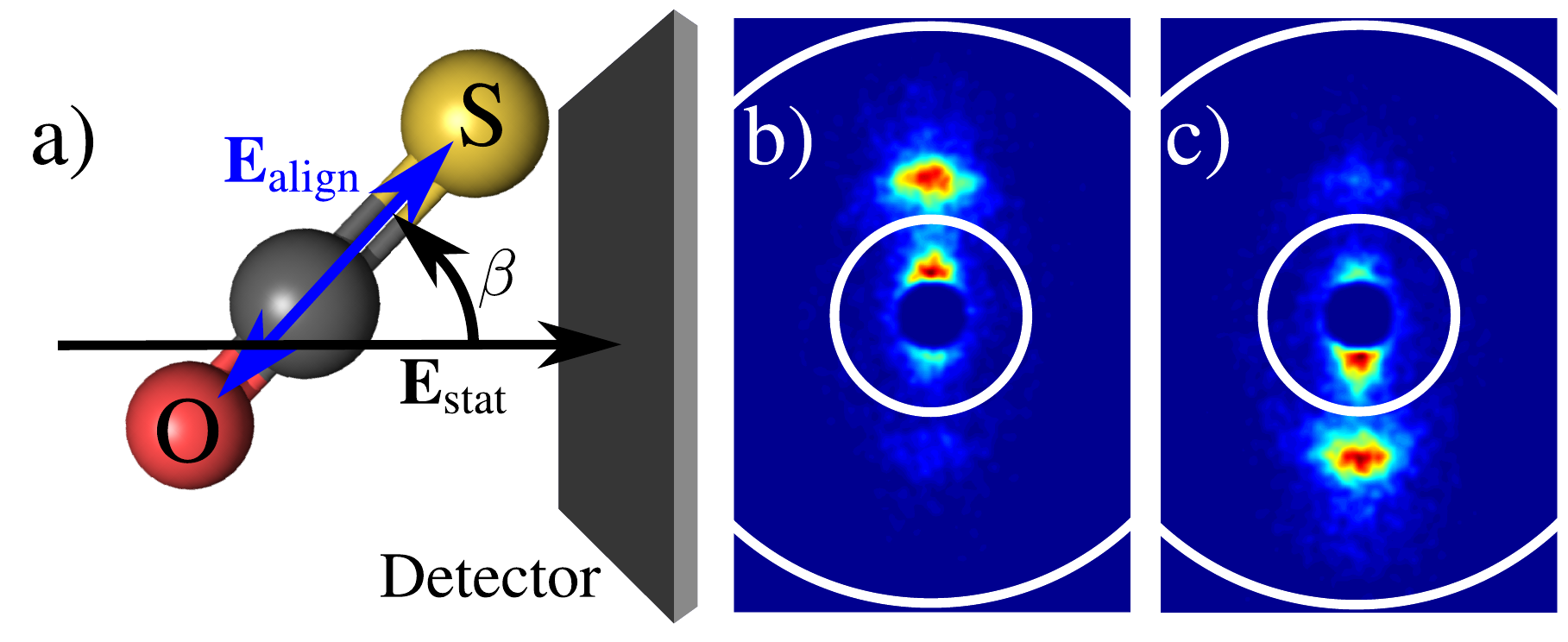}
   \caption{(Color online) (a) Schematic of the field configurations showing the polarization
      directions of the alignment and the probe pulses as well as the static field inside the VMI
      spectrometer and the definition of the angle $\beta$. (b),(c) S$^+$ ion images for
      $\beta=45\degree$ and $135\degree$, respectively and
      $\Estatabs=\SI{571}{\volt\per\centi\meter}$. The rings in the image depict the limits for
      $\text{S}^+$ ions from the $\text{S}^++\text{CO}^+$ channel used in the calculation of the
      degree of orientation. }
   \label{fig:setup}
\end{figure}

The experimental setup has been described in detail before \cite{filsinger_quantum-state_2009,
   Hansen:PRA:2011, Nielsen:PCCP13:18971} and only a few important details will be pointed out, see
\autoref{fig:setup}(a).
A pulsed molecular beam is formed by expanding a mixture of 1~mbar of OCS and 10~bar of neon into
vacuum through a pulsed valve. The molecular beam is skimmed twice before entering a 15-cm-long
electrostatic deflector. Here it is spatially dispersed in the vertical direction according to the
effective dipole moments of the quantum states~\cite{filsinger_quantum-state_2009}. Hereafter, the
molecules travel into a velocity map imaging (VMI) spectrometer where they are crossed by two pulsed
laser beams. The first pulse ($\Ealign$, $\lambda=\SI{1064}{\nano\meter}$,
$\tau_\text{FWHM}=\SI{8}{\nano\second}$, linearly polarized) provides the laser field for the
mixed-field orientation whereas the weak static field, $\Estat$, exploited for the orientation is
(inherently) provided as part of the VMI spectrometer, which also defines its direction. The second pulse (probe,
$\lambda=\SI{800}{\nano\meter}$, $\tau_\text{FWHM}=\SI{30}{\femto\second}$, linearly polarized) is
used to characterize the orientation and alignment by multiply ionizing the molecules, this is
followed by Coulomb explosion and imaging of the recoiling S$^+$ fragments on a two-dimensional
detector.

The strongest orientation is expected when $\Ealign$ is parallel to $\Estat$. This geometry is,
however, not well suited for the ion imaging method to characterize the orientation because the
experimental observable, the S$^+$ ions, will then be localized in the center of the detector.
Consequently, all measurements are conducted with $\Ealign$ rotated by an angle $\beta\ne0$ with
respect to $\Estat$ (see \autoref{fig:setup}(a)). Figures~\ref{fig:setup}(b) and (c) show examples
of S$^+$ ion images for $\beta=45\degree$ and $135\degree$ (equal to $-45\degree$). The S$^+$ ions
from the Coulomb explosion channel S$^++\text{CO}^+$, appearing in the outermost part of the images,
are highly directional and provide direct information about the alignment and orientation of the OCS
molecules at the time of ionization.

The strong angular confinement of the S$^+$ ions shows that the OCS molecules are sharply
one-dimensionally aligned along $\Ealign$. In addition, a pronounced asymmetry of the S$^{+}$ ions
emitted either along or opposite to $\Estat$, with an excess of S$^+$ in the upper (lower) region
for $\beta=45\degree~(135\degree)$ is observed. This shows that the molecules are oriented with the
S-end preferentially pointing toward the detector screen -- as expected and in agreement with
previous studies~\cite{Dimitrovski:PRA:2011}. To quantify the degree of orientation only ions from
the $\text{S}^++\text{CO}^+$ channel are considered. We then specify the orientation by the ratio
\Nuptotal{} of the number of these ions in the upper half of the image $\text{N}_\text{up}$ compared
to the total number of ions $\text{N}_\text{tot}$, from this channel.

\begin{figure}
   \includegraphics[width=\figwidth]{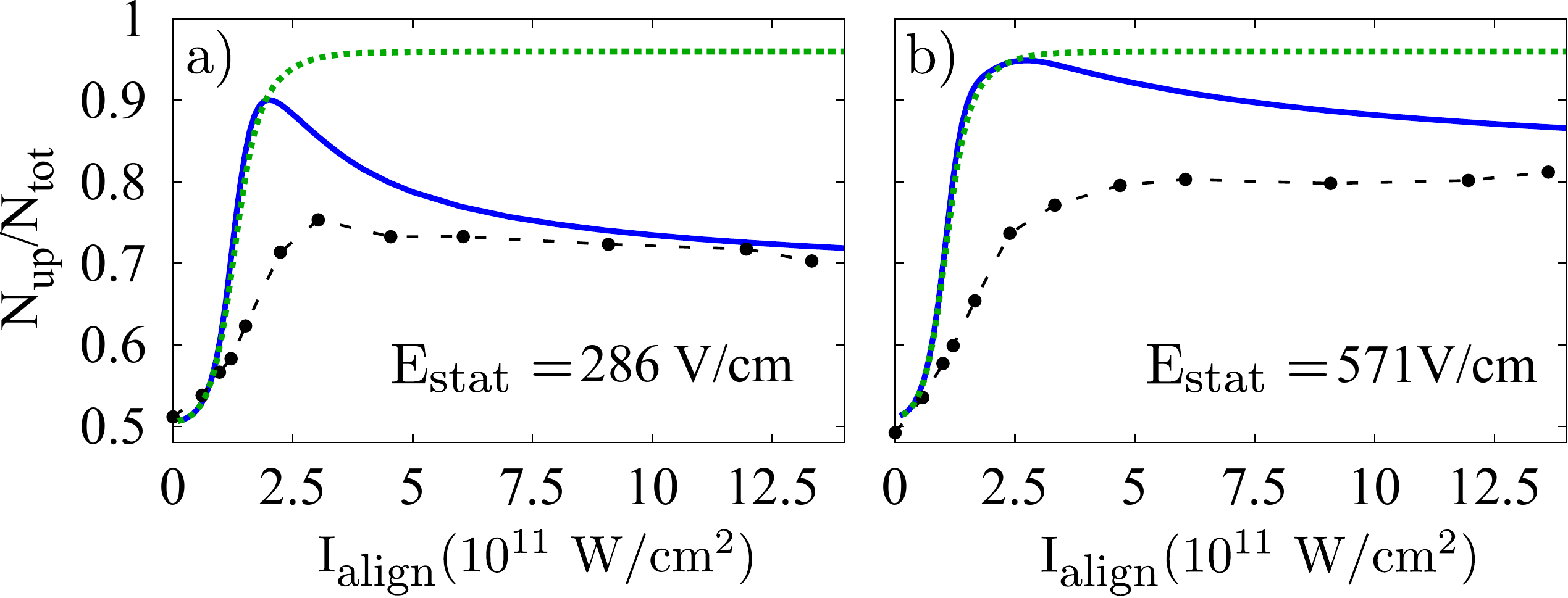}
   \caption{(Color online) Orientation ratio for $\beta=45\degree$ as a function of $\Ialign$, for
      the weak (a) and the strong (b) static field, showing the experimental results (black
      circles), the adiabatic calculations (green dotted line) and the time-dependent calculations
      (blue solid line).}
   \label{fig:ori_intensity}
\end{figure}

In the first set of measurements the degree of orientation is recorded as a function of the
alignment pulse intensity, $\Ialign$, for two values of $\Estatabs$. The results are shown in
\autoref{fig:ori_intensity}.
For low values of $\Ialign$ the orientation ratio is almost the same for the two static fields but
for $\Ialign>\SI{2.5e11}{\intensity}$ the results differ. For the strong static field the
orientation reaches a maximum of approximately 0.8 at $\Ialign=\SI{5e11}{\intensity}$ and remains
essentially constant out to \SI{1.4e12}{\intensity}. In contrast, for the smaller static field the
maximum orientation occurs already at $\Ialign=\SI{3e11}{\intensity}$ and the degree of orientation
decreases as $\Ialign$ is further increased, dropping to 0.70 at $\Ialign=\SI{1.4e12}{\intensity}$.
The calculated degree of mixed-field orientation using the adiabatic model for $\beta=45\degree$, as
in the experiment, is shown in \autoref{fig:ori_intensity}. Here, we have used rotational-state
populations, in the coordinate system of the electric field in the deflector, with $92$~\% in the
\dstate{0}{0} state, adiabatically corresponding to the field-free $J=M=0$ state, \SI{4}{\percent}
in the \dstate{1}{1} state and \SI{4}{\percent} in the \dstate{1}{-1}
state~\cite{Nielsen:PCCP13:18971}. These states are projected onto a coordinate system for the
mixed-field orientation that is defined by $\Ealign$. The properly symmetrized states are then
\pstate{0}{0}{e}, \pstate{1}{1}{e}, and \pstate{1}{1}{o}, where $e$ and $o$ denote even and odd
parity with respect to the plane defined by \Estat\ and \Ealign, respectively. The volume
effect~\cite{omiste-pccp-2011} is accounted for by using an experimentally determined cubic
dependence on the probe pulse intensity. These calculations predict that the orientation is
independent of the applied static field. Following a rapid initial rise it reaches a value of 0.96
already at \SI{4e11}{\intensity} and remains constant. Clearly, these predictions are at odds with
the experimental findings: The simulated degree of orientation is much too strong and it does not
reproduce the decrease of the orientation at intensities above \SI{3e11}{\intensity}, that is
experimentally observed for the case of the smaller static field.

\begin{figure}
   \includegraphics[width=\figwidth]{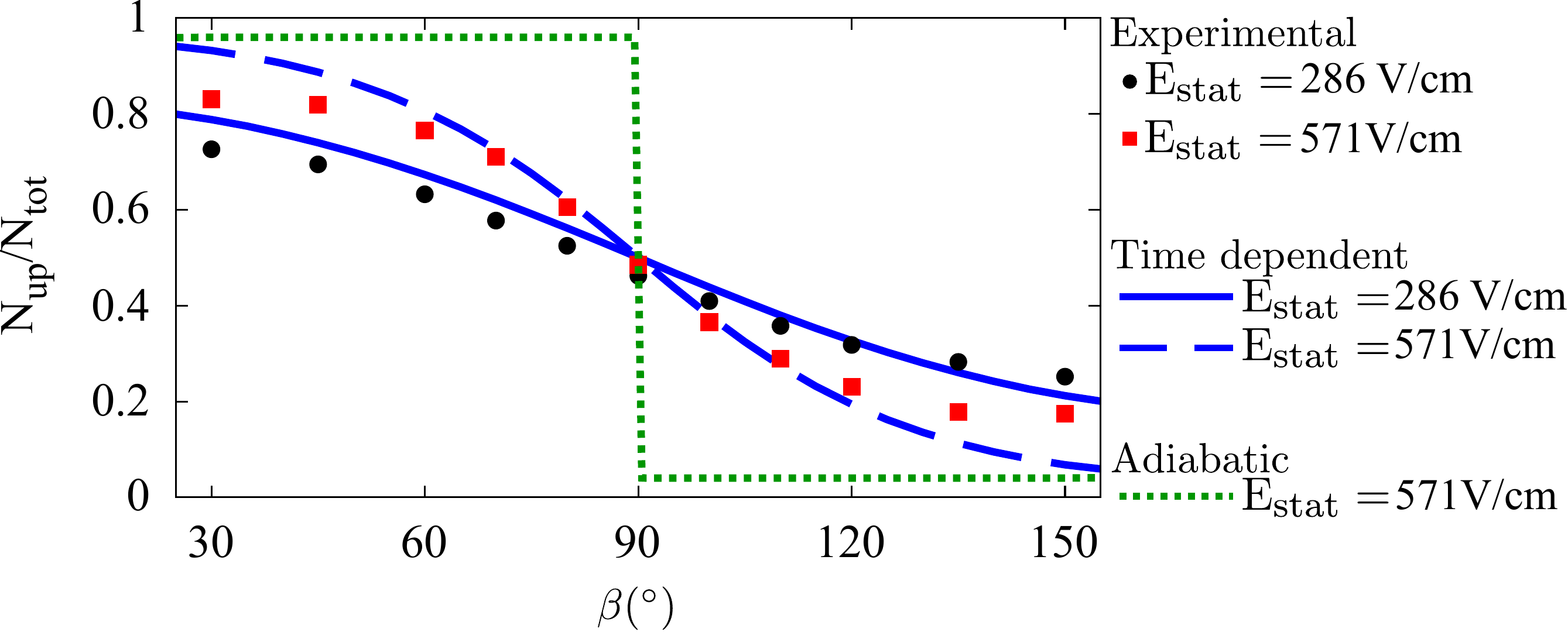}
   \caption{(Color online) The orientation ratio as a function of $\beta$ for
      $\Ialign=\SI{9.1e11}{\intensity}$ for the experiment using
      $\Estatabs=\SI{286}{\fieldstrength}$ (black circles) and $\Estat=\SI{571}{\fieldstrength}$
      (red squares), for the adiabatic calculations, which are identical for the two static fields
      (green dotted line), and for the time-dependent calculations using
      $\Estatabs=\SI{286}{\fieldstrength}$ (blue dashed line) and
      $\Estatabs=\SI{571}{\fieldstrength}$ (blue solid line).}
   \label{fig:ori_beta}
\end{figure}
In the second set of measurements, shown in \autoref{fig:ori_beta}, the degree of orientation is
recorded as a function of $\beta$ for the weak and the strong static fields and for a fixed value of
$\Ialign$ of \SI{9.1e11}{\intensity}. For both static field strengths $\Nuptotal$ decreases
monotonically as $\beta$ increases from $30\degree$ to $150\degree$. At all $\beta$ values the
strong field leads to stronger orientation than the weak field. The $\Nuptotal$ ratios calculated
from the adiabatic model are essentially identical for the two static field strengths. The sharp
rise (fall) of the curve to a value close to 0.96 (0.04) as $\beta$ is increased (decreased) below
(above) $90\degree$ shows that very strong orientation is reached already for a very modest static
electric field along $\Ealign$. This calculated behavior of the orientation differs qualitatively as
well as quantitatively from the experimental results.

To obtain a better model of the mixed-field orientation process we solve the time-dependent
Schr\"odinger equation using the experimental field configurations and rotational-state
populations~\cite{sanchezmoreno:PRA.2007}. The results for $\Nuptotal$ as a function of $\Ialign$
are shown in \autoref{fig:ori_intensity}. The predictions of stronger orientation for the strong
static field and, in the weak static field case, the decreasing orientation at increasing intensity
for $\Ialign>\SI{3e11}{\intensity}$ are in line with the experimental findings. Moreover, the smooth
$\beta$-dependence of the orientation, shown in \autoref{fig:ori_beta}, is fully captured by the
time-dependent calculations. Quantitatively, the calculated values overestimate the degree of
orientation. This could partly be due to temporal substructure in the experimentally applied laser
pulses, which could induce more nonadiabatic population transfer. Overall, these non-adiabatic-model
calculations are in much better agreement with the experimental results than the adiabatically
calculated ones.

\begin{figure}
   \includegraphics[width=\figwidth]{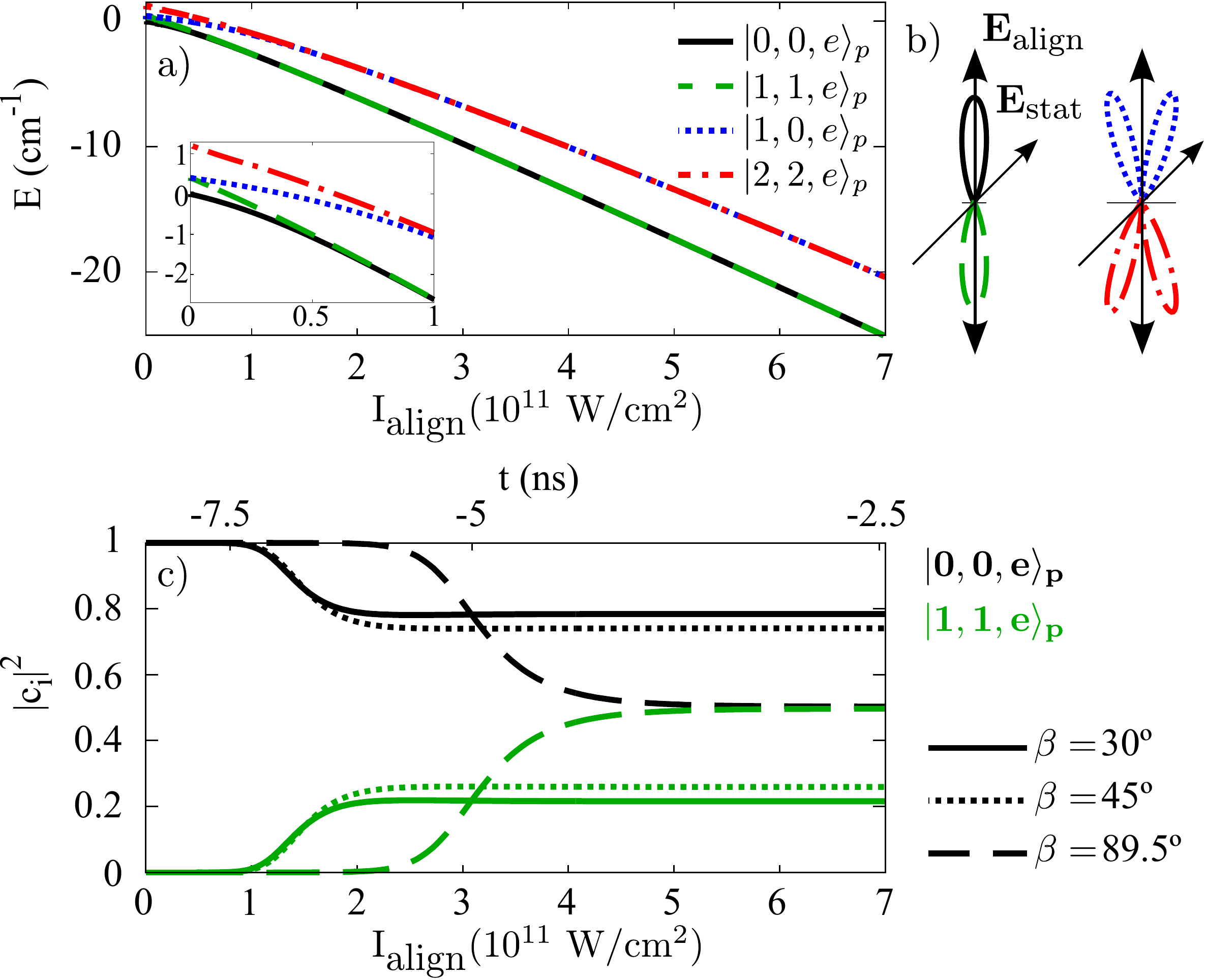}
   \caption{(Color online) (a) Energy of the 4 lowest lying rotational eigenstates as a function
      $\Ialign$ (time). The inset shows the relevant energy and intensity ranges where the formation
      of the near-degenerate doublets occurs. (b) Polar plot representation of the wavefunctions for
      the 4 states shown in (a) at $\Ialign=\SI{9e11}{\intensity}$. The single-headed arrow shows
      the direction of the static field. (c) The squares of the coefficients for the projection of
      the time-dependent pendular wavefunction of the absolute ground state onto the adiabatic
      pendular state basis (\ppstate{0}{0}{e} (black) and \ppstate{1}{1}{e} (green)) as a function
      of $\Ialign$ (time) for $\beta=30\degree$ (solid line), $45\degree$ (dotted line) and
      $89.5\degree$ (dashed line). $\Estatabs=\SI{286}{\fieldstrength}$ for all data.}
   \label{fig:starkshift}
\end{figure}

The underlying physical picture for understanding the failure of the adiabatic model is obtained by
considering the evolution of the states during the turn-on of the alignment pulse. Before the pulse,
the rotational states are essentially described by field-free rotor states. As the laser field
strength increases, the states are hybridized by the combined action of the laser and static fields.
In \autoref{fig:starkshift}(a) the formation of doublets of nearly-degenerate pendular states in the
strong laser field regime is shown.
For the laser and static fields used in the experiment the absolute ground state \ppstate{0}{0}{e}
is right-way oriented, \ie, the permanent dipole moment is pointing along $\Estat$. The upper level,
\ppstate{1}{1}{e}, of the lowest doublet is wrong-way oriented - see \autoref{fig:starkshift}(b).

As the alignment field is turned on the states that eventually form the near-degenerate doublet are
coming closer together. This is illustrated in \autoref{fig:starkshift}(a) for the \pstate{0}{0}{e},
\pstate{1}{1}{e} pair and the \pstate{1}{0}{e}, \pstate{2}{2}{e} pair and it results from the ac
Stark interaction. When the energy splitting within a pair approaches the coupling strength due to
the dc Stark interaction between the two sublevels, the two states in each pair can mix because they
have the same symmetry, provided the laser and static fields are non-perpendicular. This will result
in population transfer between the oriented and anti-oriented states. The probability for mixing,
corresponding to a crossing from one state of the doublet to the other, is determined by the rate of
the turn-on and the energy separation between the \ppstate{0}{0}{e} and \ppstate{1}{1}{e} states. If
the splitting is small, which is the case for a weak static electric field, the two states will be
strongly coupled. To ensure fully adiabatic transfer it is necessary to turn on the laser field on a
timescale slower than the inverse of the energy splitting of the near-degenerate doublet. This time
can be much longer than the rotational period of the molecule. For the lowest doublet formed in OCS
with $\Estatabs=\SI{286}{\fieldstrength}$ and $\Ialign=\SI{9.1e11}{\intensity}$ our calculations
show that the alignment pulse must be $\SI{50}{\nano \second}$ long to ensure adiabatic transfer.
The $\SI{8}{\nano \second}$ pulses used in the current experiments do not fulfill this adiabaticity
criterion although they are a hundred times longer than the rotational period of OCS - the condition
previously considered sufficient for adiabatic behavior.

The population transfer is illustrated in \autoref{fig:starkshift}(c), where the decomposition of
the time-dependent state, which starts as the rotational ground state, in terms of the pendular
states, is shown during the time interval representing the turn-on of the alignment pulse for three
$\beta$ values. The field-free ground state is not transferred adiabatically to the pendular ground
state: For $\beta=45\degree$, the final state is decomposed into \SI{74.06} {\percent}
\ppstate{0}{0}{e} and \SI{25.94}{\percent} \ppstate{1}{1}{e}. Therefore, the resulting degree of
orientation falls below that expected for a pure adiabatic transfer since the \ppstate{1}{1}{e}
state is wrong-way oriented. Similarly, other field-free rotational states are mixed with different
pendular states during the turn-on, for instance the initial state \dstate{1}{1} ends up in a
superposition of \SI{13.00}{\percent} \ppstate{0}{0}{e}, \SI{37.10}{\percent} \ppstate{1}{1}{e},
\SI{35.64}{\percent} \ppstate{1}{0}{e}, and \SI{14.26}{\percent} \ppstate{2}{2}{e}. For
$\beta=89.5\degree$, the electric field along the molecular axis is small, and the \ppstate{0}{0}{e}
and \ppstate{1}{1}{e} states contribute with \SI{50.32}{\percent} and \SI{49.68}{\percent},
respectively, to the time evolution of \dstate{0}{0}, resulting in a vanishing orientation. By
contrast, alignment is expected to remain strong since both the \ppstate{0}{0}{e} and
\ppstate{1}{1}{e} states imply tight confinement of the molecular axis along the laser field
polarization, see \autoref{fig:starkshift}(b). The experimental observations for perpendicular
fields do indeed show no orientation but strong alignment~\cite{holmegaard_phd_2010}.

In summary, the combined action of a moderately strong laser field and a weak electrostatic field
remains an attractive approach for creating tightly oriented molecules, but to fully exploit the
potential of the method it is necessary to redefine the meaning of adiabatic conditions. Unlike
alignment, where adiabaticity is ensured by turning on the laser field slower than the rotational
period of the pertinent molecule, adiabatic transfer in orientation necessitates that the laser
field be turned on slower than the inverse of the minimum spacing between the two pendular states in
a doublet. This has repercussions for designing experimental parameters such that the degree of
orientation be optimized. In the case of OCS, our calculations show that when
$\Estatabs=\SI{286}{\fieldstrength}$ and $\beta=45\degree$ adiabatic transfer of the \dstate{0}{0}
state to the \ppstate{0}{0}{e} state is obtained if a transform-limited laser pulse with a Gaussian
pulse duration (full width half maximum) of $\SI{50}{\nano\second}$ is used. In practice, such
pulses are not easily supplied by lasers typically present in laboratories. Considering instead the
\SI{10} {\nano \second} output from the widespread Nd:YAG lasers adiabatic transfer will occur for a
static electric field of $\SI{2}{k\volt\per\centi\meter}$: The increased static field leads to a
larger minimum spacing of the doublet and, thus, relaxes the requirement for the slowness of the
laser-field turn-on. Such static fields are compatible with, for instance, VMI spectrometers. For a
pulse durations of $\SI{500}{\pico \second}$, which would be relevant for using the stretched output
from amplified Ti-Sapphire lasers, a static field of $\SI{10}{k\volt\per\centi\meter}$ is needed to
ensure adiabatic conditions. Again this is compatible with electron or ion spectrometers
\cite{ghafur_impulsive_2009}.

We note that the lack of adiabatic behavior will also be influenced by avoided crossings between
rotationally excited states at low laser intensities. For larger molecules, where the rotational
level structure is quite complex, this effect is expected to be particularly important
\cite{Bulthuis:JPCA101:7684, omiste-pccp-2011}, but an increase of the static field strength should
enhance the degree of orientation as already demonstrated experimentally for several asymmetric top
rotors~\cite{Holmegaard:PRL102:023001, filsinger_quantum-state_2009, Holmegaard:NatPhys6:428,
   Hansen:PRA:2011}.

Financial support by the Spanish project FIS2011-24540 (MICINN) as well as the Grant FQM-4643 (Junta
de Andaluc\'{\i}a) is gratefully appreciated. J.\,J.\,O.\ acknowledges the support of ME under the
program FPU. R.\,G.\,F.\ and J.\,J.\,O.\ belong to the Andalusian research group FQM-207.

\bibliographystyle{jk-apsrev}
\bibliography{OCS-orientation}
\end{document}